\documentclass[aps,pra,preprint,superscriptaddress,longbibliography]{revtex4-1}

\usepackage{amsmath}
\usepackage{graphicx}
\usepackage{times}
\usepackage{amssymb}
\usepackage{hyperref}
\usepackage{textcomp}
\usepackage{xcolor}

\begin{document}

	\title[]{Radiation pressure in finite Fabry-P\'erot cavities}

	\author{P.~Gr\"unwald}
	\affiliation{
		Tecnologico de Monterrey, Escuela de Ingenier\'ia y Ciencias, Ave. Eugenio Garza Sada 2501, Monterrey, N.L., M\'exico, 64849.}
	\email[email: ]{peter.gruenwald@tec.mx}

	\author{B.~M.~Rodr\'iguez-Lara}
	\affiliation{
		Tecnologico de Monterrey, Escuela de Ingenier\'ia y Ciencias, Ave. Eugenio Garza Sada 2501, Monterrey, N.L., M\'exico, 64849.}
	\affiliation{Instituto Nacional de Astrof\'{\i}sica, \'Optica y Electr\'onica, Calle Luis Enrique Erro No. 1, Sta. Ma. Tonantzintla, Pue. CP 72840, M\'exico.}

\begin{abstract}
	
We study the effect of finite size and misalignment on a fundamental optomechanical setup: a Fabry-Pérot cavity with one fixed and one moveable mirror.
We describe in detail light confinement under these real world imperfections and compare the behaviour of the intracavity and output fields to the well-known ideal case.
In particular, we show that it is possible to trace the motion of the movable mirror itself by measuring intensity changes in the output field even in the presence of fabrication shortcomings and thermal noise. 
Our result might be relevant to the transition from high precision research experiments to everyday commercial applications of optomechanics; such as high-precission stepmotor or actuator positioning.

	\end{abstract}

	\maketitle

\section{Introduction}
Optomechanics (OM) studies the interaction of an electromagnetic radiation field with a mechanical oscillator. 
The dynamics of each subsystem, well-described within the classical and quantum frameworks, become non-linearly coupled, revealing a plethora of new fundamental effects as well as applications \cite{Milburn2011,Meystre2013,Aspelmeyer2014,Milburn2016}. 
OM systems range from macroscopic mirrors used to detect gravitational waves \cite{GravWave01,GravWave02} to microscopic cantilevers\cite{Gigan2006,Arcizet2006,Kleckner2006} and membranes \cite{Thompson2008} used in the search for macroscopic quantum state engineering.
Applying a well-defined driving field, the state of the mechanical oscillator can be prepared, and then coherently controlled~\cite{Coh01,Coh02,Coh03}. 
This allows, for example, to either cool the mechanical motion or amplify the forces acting on it solely by manipulating the electromagnetic field~\cite{Metzger2008}.
This is the basis for all kinds of high-precision measurements or quantum information processes. 

Radiation pressure is the origin behind optomechanical coupling \cite{Lebedev1901,Nichols1901}.
The first experimental proof of such an effect dates back to 1967 \cite{first}, while the interaction between visible light and a macroscopic mechanical oscillator was demonstrated in 1983 \cite{second}. 
Law provided the first consistent second quantization of the model in 1995 \cite{Law1995}, showing that the cavity frequency change leads, in general, to a complex coupling between creation and annihilation operators of the optical and mechanical subsystems. 
This coupling is usually restricted to the linear regime in quantum OM due to the limited amplitude of the mechanical oscillation \cite{Schliesser2010}. However, recently technical developments has put the nonlinear range of interaction within reach and jump-started research into corrections to the standard OM description \cite{Sala2017}.

We focus on a resonator that is finite in size and possible misalignment of the incident beam. 
These imperfections deterministically limit the confinement time of the field inside the resonator. 
This is an ever-present issue that state-of-the-art laboratories can suppress as much as possible for high-end technology applications.
In contrast, our motivation sparks from the potential application of these systems in everyday, mass-produced applications where an adequate characterization of imperfections might be crucial for their use.
Our setup is a driven high-finesse Fabry-Perot (FP) resonator made from two plane, highly-reflecting mirrors \cite{FabryPerot}.
We consider the classical optical response function of the cavity~\cite{second,Gozzini1985,Ujihara1985} driven by a monochromatic laser. 
Radiation pressure produces a displacement of the movable mirror that, in turn, leads to a change in the phase that the intracavity field accumulates and, in the end, modifies the sensitivity of the interferometer. 
In general, there is a trade-off between that sensitivity and the broadness of the FP radiation pressure (FPRP) resonance for varying the reflectivities of the mirrors. 
This trade-off is measurable in the output field in front of the fixed mirror. 
In this case, however, we show that the effect of the reflectivity of the mirrors on the output intensity is suppressed, allowing a stable measurement. 

In the following, we review the results for an ideal cavity for the sake of comparison. 
Then, we incorporate the effect of finite cavity size and input beam misalignment to show that the intracavity response function for limited confinement time is just the ideal one times a fast oscillating factor controlled by the number of intracavity reflections. 
We also discuss the effect of limited mirror reflectivity; it causes further broadening and scaling-down of the FPRP resonance around its maximum value but the phase sensitivity and the output field behaviour is mostly preserved.
Afterwards, we analyse how these finite size effects would reflect on precision measurements including thermal mechanical noise. 
Finally, we provide some conclusions and an outlook.

\section{Ideal Fabry-P\'erot Cavity Review}\label{sec.Infinite}
%\subsection{Response function}

\begin{figure}[h]
	\begin{center}
		\includegraphics[scale=1]{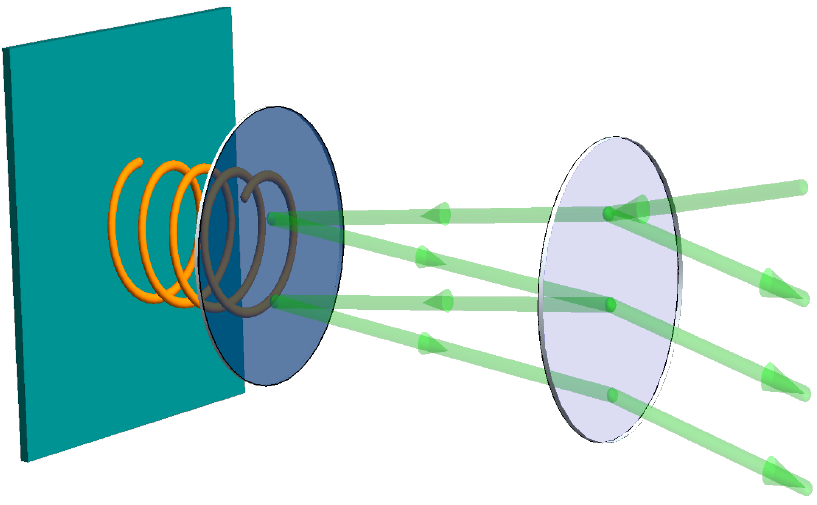}
	\end{center}
	\caption{Schematic for a cavity with one moveable (left) and one fixed (right) mirror. Laser driving light impinges on the cavity at a given angle, akin to a FP resonator, creating multiple but finite reflections on the mirrors.}\label{fig.setup}
\end{figure}

Let us first recap the well-known results for the radiation-pressure-induced response of the optomechanical system depicted in Fig.~\ref{fig.setup} \cite{Sala2017}. A cavity made of two highly-reflective mirrors, one allowed to undergo harmonic motion and the other fixed, separated by a distance $L$ can be excited to a certain resonance, $\omega_0 = n \pi c\cos \theta / L$ with $n \geq 1$ the number of the excited harmonic and $\theta$ the angle between the normal of the cavity walls and the direction of the incident driving monochromatic laser of frequency $\omega_L$~\cite{Gong2009}.
Radiation pressure arises from momentum transfer of the light field to the moving mirror and is proportional to the Poynting vector $|S|\propto |E|^2$, with $E$ the total electric field impinging on the moveable mirror. We assume the transverse area of the laser constant as the laser travels through the cavity. 
The total field amplitude $E$ can be given in terms of the response function $J(\varphi)=|E/E_{\textrm L}|^2$ that relates it with respect to the driving laser intensity $E_{L}$.
For an infinitely long cavity, we can calculate the transfer function,
\begin{eqnarray}
J(\varphi)
&=& J_{\mathrm{max}} A(\varphi),
\end{eqnarray}
in terms of the ideal maximum response,
\begin{equation}
J_{\mathrm{max}}=\frac{1-R_{\textrm f}}{\left(1  - \sqrt{R_{\textrm f} R_{\textrm m}}\right)^2}, \label{eq:Jmax}
\end{equation}
obtained at the half-round trip phase $\varphi = n \pi$, the Airy function, and the coefficient of finesse \cite{FabryPerot},
\begin{equation}
A(\varphi)=\frac{1}{1+F\sin^2(\varphi)},\quad F=\frac{4\sqrt{R_\textrm fR_\textrm m}}{\left(1-\sqrt{R_\textrm fR_\textrm m}\right)^2},
\end{equation}
in that order.
Here the reflectivities of the moveable and fixed mirrors are given by the parameters $R_{\textrm m}$ and $R_{\textrm f}$, respectively.
The  half-round trip phase gained by the field inside the cavity,
\begin{equation}
\varphi=\frac{\omega_\textrm L}{c}\cdot\frac{L+x}{\cos\theta}=n\pi\frac{\omega_\textrm L}{\omega_0}\left(1+\frac{x}{L}\right)=\varphi_0\left(1+\frac{x}{L}\right),
\end{equation}
is proportional to the fixed resonator half-round trip phase $\varphi_{0}$ and the position of the moveable mirror $x$.

We find it useful to consider two particular cases of mirror configurations: (I) one where the mirrors are identical, $R_\textrm f=R_\textrm m=R$, and (II) another where the moveable mirror is ideal, $R_{\textrm f} =  R < R_{\textrm m} =1$. 
In the first case, we find that the response function changes from $(1-R)^{-1} \gg 1$ to approximately $(1-R)/4$.  
In the second, it changes from $(1+\sqrt R)/(1-\sqrt R)$ to $(1-\sqrt R)/(1+\sqrt R)$. 
For a high reflectivity of $R=0.999$, this means a change of $4\times 10^6$ for case (I) and $1.6\times 10^7$ for case (II). 
Such an enormous change in the response function leads to the question of phase sensitivity. 
The phase that provides us with $J(\varphi)=1$ will be denoted $\varphi_\textrm{eq}$ and is given by 
\begin{eqnarray}
\sin(\varphi_\textrm{eq})&=&\sqrt{\frac{1}{2}\left[1-\sqrt{R_\textrm f}\left(\frac{\sqrt{R_\textrm m}+\sqrt{R_\textrm m}^{-1}}{2}\right)\right]}.
\end{eqnarray}
At this point the response of the resonator becomes as small as if the laser light would directly impinge just once on the moveable mirror.
For our two cases, this phase reduces to
\begin{eqnarray}
\sin(\varphi_\textrm{eq,I})&=&\sqrt{\frac{1-R}{4}},\quad \sin(\varphi_\textrm{eq,II})=\sqrt{\frac{1-\sqrt R}{2}},\label{eq.phi_equal}
\end{eqnarray}
and we find $\varphi_\textrm{eq,I}\approx\varphi_\textrm{eq,II}\approx0.906^\circ$ for a reflectivity value of $R=0.999$  . 
We see that even the tiniest deviation from resonance drastically diminishes the effect of FPRP. 
For both cases, if the following condition holds,
\begin{equation}
R_\textrm f>\frac{4R_\textrm m}{\left(1+R_\textrm m \right)^2}\geq R_\textrm m,
\end{equation}
there is no real solution for $\varphi_{eq}$, as $J(\varphi)<1$ for all values of $\varphi$. It is thus paramount to have the reflectivity of the moveable mirror at least as large as, if not significantly larger than, the reflectivity of the fixed mirror.

\begin{figure}[h]
	\includegraphics[scale=1]{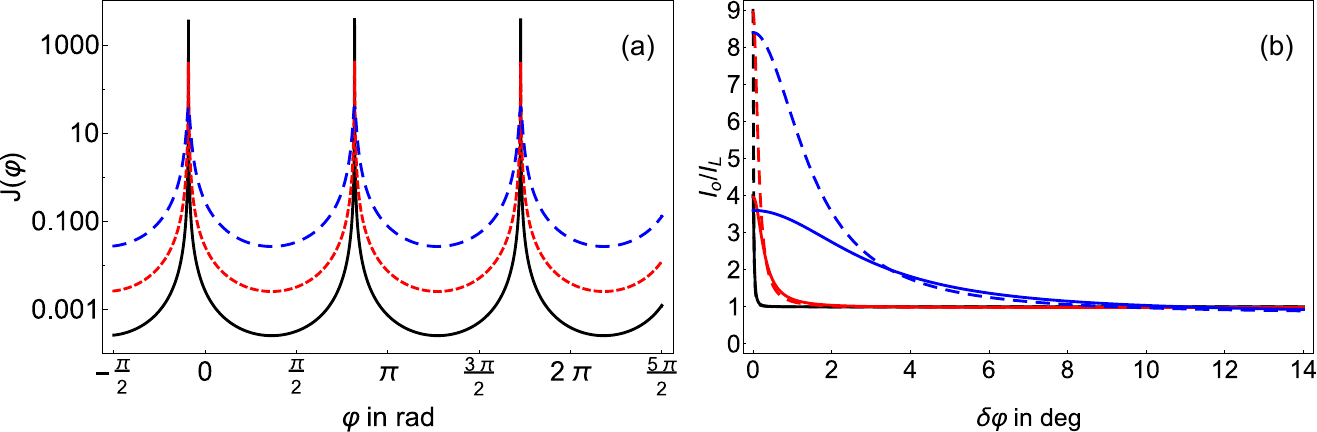}
	\caption{(Color online) (a) ideal response function, $J(\varphi_{s})$, in terms of the fixed half-round trip phase, $\varphi_0$, for $n=1$ in case II with $\varphi_{s} =\varphi_0(1+x_\textrm s/L)$ and  $x_\textrm s/L=0.1$. The different colors represent $R=0.999$ (solid black), $0.99$ (dashed blue), $0.9$ (wider dashed red). (b): Output intensity $I_\textrm o/I_\textrm L$ as a function of the deviation $\delta\varphi$ from the FPRP resonance $n\pi$ for case (I) (solid lines) and (II) (dashed lines) with $R=0.999$ (black), $R=0.99$ (red), $R=0.9$ (blue).}\label{fig:IdealResponse}
\end{figure}

Here, we can take a stop and make two assumptions: (i) the driving laser is dominant, thus the field quadratures follow only this laser oscillation in the long-time limit, and (ii) the position of the moveable mirror can be approximated by a constant in the long-time limit, $\lim_{t \to \infty} x \approx x_{\textrm s}$, due to the absence of direct driving and the presence of damping. This steady state can either be numerically evaluated from the balance of forces, or measured indirectly. These assumptions yield for the long-time limit of the half-round trip phase
\begin{equation}
\varphi_{s}=\lim\limits_{t\to\infty}\varphi=\varphi_0\left(1+\frac{x_\textrm s}{L}\right).
\end{equation}
The driving laser and cavity field pressure contributions provide only positive moveable mirror displacements, $x_\textrm s>0$, for all cases. 
Thus, we are required to set the laser below the bare resonance, $\omega_{\textrm L} < \omega_{0}$, in order to reach the maximum of the response function, $J_{\mathrm{max}}$, in the long-time limit.
The response functions for three different reflectivities in case (II) are depicted in Fig.~\ref{fig:IdealResponse}(a). It is clear that there is a tradeoff between a large intensity jump in the response function, and its phase broadness around $\varphi=n\pi$. That means that for high reflectivities, small amounts of shift of the moveable mirror may lead to big variations in the response function. Let us reconsider Eq.~(\ref{eq.phi_equal}) for case (II) and express the phase shift $\delta\varphi$ from FPRP resonance in terms of the cavity wavelength $\lambda$ corresponding to $\omega_0$. We find
\begin{equation}
	\delta\varphi=\varphi-\varphi_0=n\pi\frac{\omega_\textrm L}{\omega_0}\frac{x_\textrm s}{L}=\frac{2\pi\omega_\textrm L}{\omega_0\cos\theta}\frac{x_\textrm s}{\lambda}.
\end{equation}
For the laser being on bare cavity resonance and almost in normal incidence ($\theta\approx0$), the mirror shift compared to the wavelength is just $\delta\varphi/(2\pi)$. For a high reflectivity $R=0.999$ a phase-shift to $\varphi_\textrm{eq}$ thus corresponds to a movement of $x_\textrm s\approx0.25\%\lambda$. For a He-Ne-laser driven cavity this equals 1.6~nm. Within a movement of the mirror of 1.6~nm, the response function changes by a factor of 4000. That is, if we can experimenteally determine this quantity we would be extremely sensitive in a very small region. On the other hand, for $R=0.9$ and case (II), the response changes only by a factor of roughly 38, but this change occurs over a range of $x_\textrm s\approx2.56\%\lambda$ or 16~nm.

The actual displacement of the moveable mirror $x_\textrm s$ is generally hard to measure. Thus, we analyze the output intensity $I_\textrm o\propto|E_\textrm o|^2$, transmitted back through the fixed mirror, instead. Making a similar calculation as for the field hitting the moveable mirror we obtain
\begin{eqnarray}
	\frac{I_\textrm o}{I_\textrm L}=\left|\frac{E_\textrm o}{E_\textrm L}\right|^2&=&R_\textrm f+[R_\textrm m(1-3R_\textrm f)+%\nonumber\\
	2\sqrt{R_\textrm fR_\textrm m}\cos(2\varphi)]J(\varphi).\label{eq.transrel}
\end{eqnarray}
The first term on the right-hand side of Eq.~(\ref{eq.transrel}), $R_\textrm f$, stems from the initial reflection when entering the cavity. 
The left-hand side would just be this $R_\textrm f$ without a second mirror, $R_\textrm m\to0$, implying that a deviation from that value is fully based on the constructive interference displayed by the response function.  Both cases (I) and (II) for different reflectivities are depicted in Fig.~\ref{fig:IdealResponse}(b).

The maximum value, given for FPRP resonance in both cases, can be shown to be
\begin{equation}
	\frac{I_\textrm o}{I_\textrm L}=4R\lesssim 4.
\end{equation}
for case (I) and 
\begin{equation}
	\frac{I_\textrm o}{I_\textrm L}=(1+2\sqrt{R})^2\lesssim 9
\end{equation} 
for case (II). In contrast to the response function $J(\varphi)$ this maximum is little effected by variations of $R\lesssim 1$, making this maximum resistent to the actual mirror parameters. For large reflectivities the phase dependence of the output field mimics the behaviour of the reflected field, just for a smaller parameter region. Thus, we only see a sharp peak around $\delta\varphi=0$, whereas for all other phases the intensity is almost constant. If the reflictivities go down on the other hand, we again see this tradeoff between lower intensity change and broader range of phase deviations for which this change occurs.
That means higher reflectivities should allow an extremely sensitive measurement of the mirror movement, but only for a very small range of movements. On the other hand lower reflectivities are less sensitive, but on a broader range. 

Quantitatively we can use the results for the response function again, because, one can easily show that
\begin{equation}
	\frac{I_\textrm o}{I_\textrm L}(\varphi_\textrm{eq})=R(3-2R)\approx 1
\end{equation}
for case (I) and 
\begin{equation}
	\frac{I_\textrm o}{I_\textrm L}(\varphi_\textrm{eq})=1
\end{equation} 
for case (II). In other words, in case (I) and for large reflectivity $R=0.999$, the output intensity changes from $4I_\textrm L$ to $I_\textrm L$ when the moveable mirror position $x_\textrm s$ changes by $0.25$~\% of the cavity wavelength.

\section{Finite Size effects}\label{sec.Finite}

The finite length of planar mirror walls implies that light at non-normal incidence, $\theta\neq0$, will only stay a limited amount of time inside the cavity, independent of the reflectivities. In this case, the geometric series at the heart of the response function becomes limited, yielding just a scaled version of the ideal response function,
\begin{equation}
	\frac{J_N(\varphi)}{J(\varphi)}=\left(1-\sqrt{R_\textrm f^NR_\textrm m^{N}}\right)^2+4\sqrt{R_\textrm f^NR_\textrm m^{N}}\sin^2(N\varphi),
\end{equation}
where $N$ is the number of hits on the back mirror. This number can be calculated from the system geometry or recovered from the maximum of the finite response 	function at $\varphi=n\pi$ for non-ideal reflectivitys $R_{\textrm{f,m}}<1$, 
\begin{equation}
	N=\frac{\ln\left[1-\sqrt{J_N(n\pi)/J(n\pi)}\right]}{\ln \sqrt{R_\textrm fR_\textrm m}}\approx\sqrt{\frac{J_N(n\pi)}{1-R_\textrm f}}.
\end{equation}
In this approximation, we used the restrictions $1-\sqrt{R_\textrm fR_\textrm m}\ll1$ and $\sqrt{J_N(n\pi)/J(n\pi)}\ll1$.	
For example, for $R_\textrm f=0.999$, we need 32 bounces to obtain $J_N(n\pi)=1$, which yields a mirror size of at least $63L\tan\theta$ for the fundamental mode $n=1$. This means mirrors must be at least $348~\mathrm{nm}$ in size for He-Ne laser light, $\lambda = 633 ~\mathrm{nm}$, at $1.00^{\circ}$ incident angle without considering the obvious diffraction issues.

	\begin{figure}[h]
		\begin{center}
			\includegraphics[scale=1]{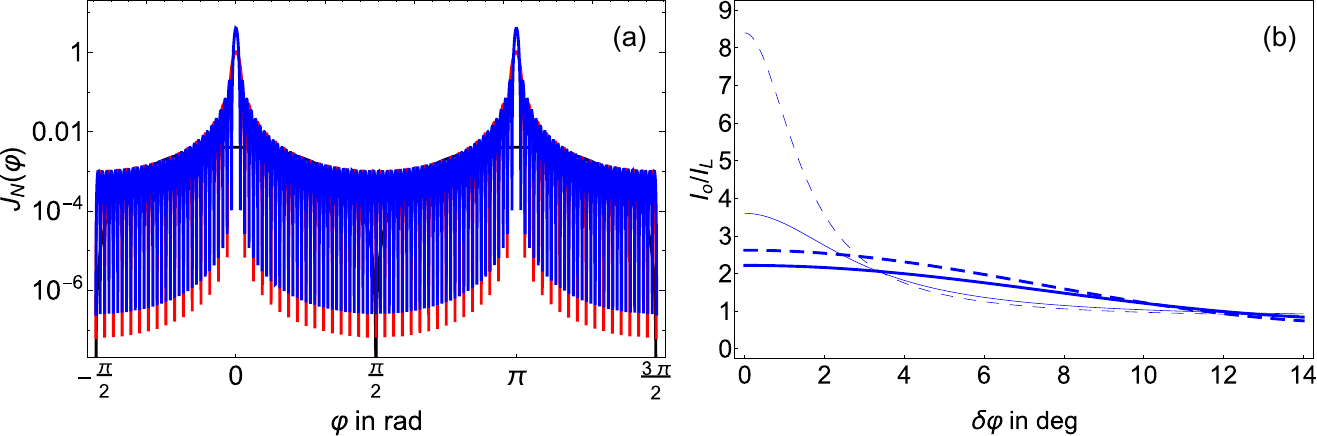}
		\end{center}
		\caption{(Color online) (a): finite response function, $J_{N}(\varphi_{s})$, for  $N=2$ (solid black), $32$ (dashed blue), and $64$ (dotted red) for case (II) and $R=0.999$. (b): Output field as in Fig.~\ref{fig:IdealResponse} for $R=0.9$ with the case $N=\infty$ depicted by the thin lines for comparison. For the thick lines $N=8$.}\label{fig:FiniteResponse}
	\end{figure}

The maxima of $J_N(\varphi)$ are broader and have lower values than the ideal case of $J(\varphi)$, and slowly approach the ideal response with increasing $N$. 
Figure \ref{fig:FiniteResponse} shows the finite response function for the cases of non-normal incidence allowing for $N=2$ bounces in solid black, where the FPRP on resonance is still smaller than if the laser light would impinge directly on the moveable mirror just once. 
The case delivering a maximum unit finite response of one, $N=32$ bounces on the moveable mirror, is shown in dashed blue and the response for double that number, $N=64$, is shown in dotted red lines yielding a maximum response slightly below to a value of four. 
The influence of the FPRP is quite fragile towards the geometry of the resonator, in particular mirror size related to incident angle and the ratio between the driving laser and the long-time cavity frequencies.

A similar but lengthy formula as Eq.~(\ref{eq.transrel}) can be derived for the output field of a finite mirror and $J_N(\varphi)$. The dependence of the output field in this case is shown in Fig.~\ref{fig:FiniteResponse}(b). In this case, we consider the low reflectivity of $R=0.9$ and only $N=8$ hits on the back mirror. 
The maximum amplitude at $\delta\varphi=0$ does not yet approach its ideal value, thus limiting the sensitivity in that range. However, in a medium range of phases, the intensity is even higher for low $N$ than for the infinite case. 
In other words, while a precise determination of the phase and thus the mirror displacement is practically impossible, it is far easier to obtain a rough estimation over a broader range of phases.

We can conclude a few things. First, the relation between input and output field at the fixed mirror is connected to the same response function $J(\varphi)$ as the reflection field at the moveable mirror. Second, the output field is limited to values below nine times the driving laser intensity and this maximum is less sensitive to the actual reflectivities than the response function. Third, for high reflectivities a very small range of phase changes, i.e. movements of the mirror, can be sensitively detected by comparing the output intensity with the laser intensity. In contrast, for lower reflectivity, a broader range of phases yields a less varying output intensity. There is a general tradeoff between the range of mirror movements that yield a variation of the output intensity and the amplitude of this variation, which we called the sensitivity. If the variation is detectable, it allows a one-to-one relation between the movement of the mirror and $I_\textrm o$. Fourth, when only a finite amount of mirror reflections is considered, the tradeoff above appears to favor a broader range while the sensitivity decreases. Additionally, for case (II), we find oscillations appearing at larger $N$, leading to unwanted ambiguities in the determination of $x_\textrm s$.

\section{Application Outlook}\label{sec.AFM}

In order to better motivate the idea that finite size effects do not destroy the applicability of these systems for precision measurements, let us consider a measurement scenario under thermal mechanical noise. 
We focus on a description via the equipartition theorem~\cite{LL3} but we note that there are other approaches to threat noise sources, for example based on the fluctuation dissipation theorem \cite{Callen1952,Liu2012}, which we do not discuss here. 
Following the standard description of a classical mechanical damped harmonic oscillator with mass $m$, frequency $f$, damping rate $\Gamma_c$, and temperature $T$, we can calculate the average thermal motion as
\begin{equation}
	\Delta x=\sqrt{\frac{k_\textrm BT}{m\Omega^2}},
\end{equation} 
with the Boltzmann constant $k_{\mathrm{B}}$.
The range of optomechanical devices and, thus, parameters for a moveable mirror is vast, so let us consider the question of mechanical noise more generally.
The output field phase is changed by the mirror movement as $\varphi=\varphi_0(1+\Delta x/L)$. 
The latter value is fixed for a given setup and temperature. 
It becomes clear that the general requirement for high precision measurement should not be larger than a fraction of a degree. 
However, due to the phase prefactor $\varphi_0\approx\pi=180^\circ$, this implies that the displacement-to-length ratio must fulfill $\Delta x/L<1/180$. 
For higher precision we must preferably aim for  $\Delta x/L \ll 1/180$.

\begin{figure}[h]
\includegraphics[scale=1]{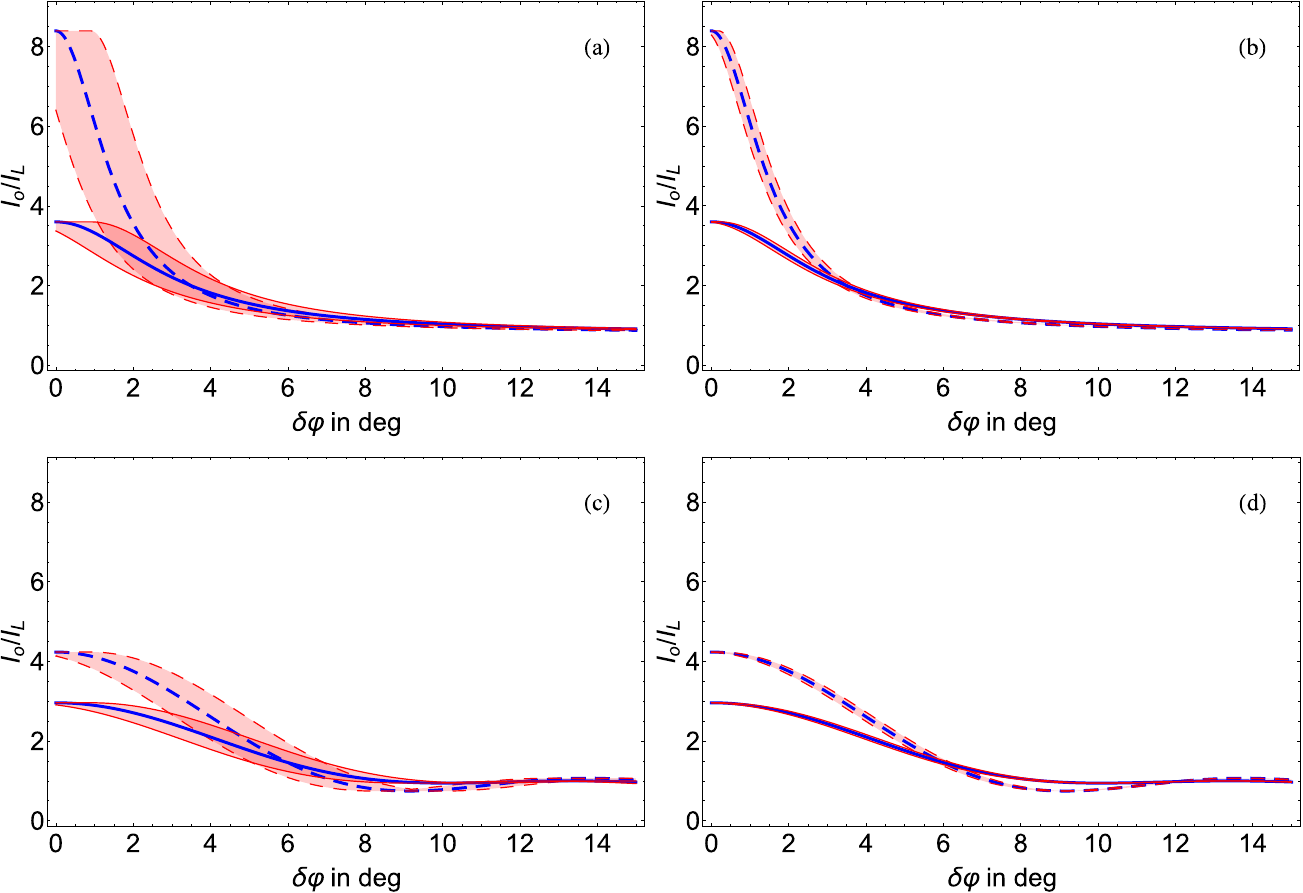}
\caption{Including mechanical thermal noise (red shaded area) to the output intensity of an infinite cavity (a),(b) for case (I) (solid) and (II) (dashed) and $R=0.9$. (a): $\Delta x/L=0.005$, (b): $\Delta x/L=0.001$. Figures (c) and (d) show the same graphs for a finite cavity with $N=16$.}\label{fig.noise1}
\end{figure}

Figure \ref{fig.noise1} depicts the effect of thermal mechanical noise on the output field intensity, normalized by the driving laser intensity, for displacement-to-length ratios of $\Delta x/L=0.005$, Fig. \ref{fig.noise1}(a) and Fig. \ref{fig.noise1}(c), as well as $ \Delta x/L=0.001$, Fig. \ref{fig.noise1}(b) and Fig. \ref{fig.noise1}(d). 
The larger the noise, the worse the sensitivity to phase changes in the output field intensity for both ideal, Fig. \ref{fig.noise1}(a) and Fig. \ref{fig.noise1}(b), and finite cavities, Fig. \ref{fig.noise1}(c) and Fig. \ref{fig.noise1}(d). The lower value of the mechanical noise is almost invisible. 
Unfortunatetly, oscillations in the output field set in for the finite case, Fig. \ref{fig.noise1}(c) and Fig. \ref{fig.noise1}(d), leading to an ambiguity above $\delta\varphi\gtrsim 6^\circ$ in the depicted examples.
Roughly speaking, one generally requires a displacement-to-length ratio $\Delta x/L<0.001$ for low noise. 
At room temperature, $T = 300$~K, this value is restricted by the condition
\begin{equation}
	m\Omega^2L^2=kL^2>~25\textrm{keV},\label{eq.energy}
\end{equation}
where the parameter $k$ is the effective spring constant of the movable mirror. 
As an example, let us consider the microelectromechanical-system cantilever with an effective spring constant 0.17~N/m from Ref.~\cite{Liu2004}. For optical driving with an He-Ne laser and a cavity with length $L=\lambda/2=316$~nm, we obtain $kL^2\approx100$~keV, which is four times as high as our lower bound. Taking into account the progress since this work our requirements should be easily realizable in everyday production.

Now, let us think about a setup where the fixed mirror is movable via a linear actuator or step motor. 
These are used to automate tiny movements in larger structures. 
Modern step motors allow a controlled periodic increment of motion by around 5~nm, not accounting for friction limitations \cite{Stepmotor}. Depending on the intensity sensitivity of our detector, we can easily reach that resolution with reflectivities of $R=0.9$ in setup (II), as the range in which strong phase sensitivity occurs is roughly 16~nm. Keep in mind that this is not the resolution limit, just the range in which the system is very sensitive.

Let us assume again the cantilever from~\cite{Liu2004}, as well as the He-Ne-laser and cavity length $L$ approximately at half the laser wavelength. Moreover, we fix the number of impinges at $N=64$. In this setup most of the ambiguities due to oscillation have died down allowing an unambiguous relation between field intensity and mirror motion on one side of the FPRP resonance. If we set $L$ exactly at half the laser wavelength, the system would be at resonance without any external laser driving. For roughly 100~mW laser power, the equilibrium point would move to the middle of the first downward slope, allowing strong phase sensitivity. Increasing the laser power further to around 200~mW, we reach the less steep region, in which a broader range of phases are covered with less change of the output intensity. If the unperturbed cavity length is slightly off from resonance, this can also be corrected by increasing or decreasing the laser power correspondingly. 
Hence, only by controlling the intensity of the driving laser we can move in or out of resonance to a point in the response function where sensitivity is lower but broader in phase change, or vice versa.
We might include such an optomechanical setup in linear actuators to switch between finder, low-precision/high-range, and measuring, high-precision/low-range, set-ups with the same device just modifying the driving laser intensity even in the presence of mechanical noise.

\section{Conclusions}\label{sec.Concl}

We studied the influence of finite size and misalignment on an optomechanical cavity setup with one moveable mirror. 
Radiation pressure pushes the cavity walls apart and creates a well-known phase-sensitive response of the intracavity field. 
For non-normal incidence of laser light, it can only impinge on the mirrors a limited number of times before leaving the cavity, independent of the mirror reflectivity. We incorporate the effects of these imperfections into the well-known dynamics of the ideal cavity setup. For highly-reflecting mirrors the evolving extreme sensitivity in a very narrow region around the resonance of the maximal response function implies that tiniest movements of the mirror yield strong variations of the field amplitude. For lower reflectivities, this sensitivity decreases but, in exchange, the phase range of the resonance increases. As a measurable quantity, we calculated the output field in front of the fixed mirror. While displaying the same sensitivity-range trade-off for measuring the phase-shift, the range of output intensities is much smaller and little affected by the actual reflectivities. Hence, similar to the response function, we can have extreme sensitivity in a very small range of mirror displacements (sub-percent of the cavity wavelength) or reasonable sensitivity in a broader range. For limited resonances and a medium number of reflections, the range of phases covered by the resonance may be even broader than for an infinite cavity. We calculated the thermal mechanical noise in such a setup to motivate using the output field to measure mirror displacements even in far-from-ideal conditions. Thermal noise appears to have a small effect for reasonable microscopic mirrors setups.

\begin{acknowledgments}
B.M.R.L acknowledges funding from Consejo Nacional de Ciencia y Tecnolog\'ia (CONACYT) (CB-2015-01-255230) and thanks Cinthia Huerta Alderete for her help with Fig. 1.
\end{acknowledgments}

%\bibliography{bib_FPRP}


\end{document}